\documentclass[11pt]{article}
\usepackage{epsfig}
\usepackage{times}

\usepackage{epsfig}
\begin{document}                
\newcommand{\keywords}{K-calculus; alternative theories; 4-dimensional optics;
 special relativity.}
\pagestyle{myheadings} \markright{K-calculus in 4-dimensional optics} 
\title{K-calculus in 4-dimensional optics}
\author{Jos\'e B. Almeida\footnote{Universidade do Minho,
Departamento de F\'isica, 4710-057 Braga, Portugal. E-mail:
\texttt{bda@fisica.uminho.pt}}}

\date{}
\maketitle
\begin{abstract}                
4-dimensional optics is based on the use 4-dimensional movement
space, resulting from the consideration of the usual 3-dimensional
coordinates complemented by proper time. The paper uses the
established K-calculus to make a parallel derivation of special
relativity and 4-dimensional optics, allowing a real possibility
of comparison between the two theories. The significance of
proper time coordinate is given special attention and its
definition is made very clear in terms of just send and receive
instants of radar pulses. The 4-dimensional optics equivalent to
relativistic Lorentz transformations is reviewed.

Special relativity and 4-dimensional optics are also compared in
terms of Lagrangian definition of worldlines and movement
Hamiltonian. The final section of the paper discusses
simultaneity through the solution of a two particle head-on
collision problem. It is shown that a very simple graphical
construction automatically solves energy and momentum
conservation when the observer is located at the collision
position. A further discussion of the representation for a
distant observer further clarifies how simultaneity is
accommodated by 4DO.

\noindent{\hrulefill}


\end{abstract}
\section{Introduction}
In recent times the author has written several papers on the
subject of his proposed alternative to general relativity,
\cite{Almeida00:4, Almeida01, Almeida01:2, Almeida01:4,
Almeida01:5} all of them rejected by the journals to which they
were submitted; refs.\ \cite{Almeida00:4} and \cite{Almeida01}
were accepted as posters in conferences. The main reason for
rejection can be found in the words of one referee:

\begin{quotation}{The foundation of the
author's model is inconsistent with SR (special relativity). In
particular, he gives no compelling reason why we should abandon
SR which is a highly successful theory. Just to turn to an
alternative speculation since, according to the author, SR is too
complicated, is no sufficient reason. Therefore I have no option
but to suggest to the editor of the \ldots\ to reject this paper.}
\end{quotation}

One should therefore ask why any effort should be made to set up
an alternative to an existing theory that has proven so
successful. There are at least two reasons to do so: In the first
place any theory is just as good as the predictions it allows for
observable phenomena and if two competing theories allow the same
predictions they are equally good until some phenomenon is better
predicted by one of them. Until one is proven better than the
other, the coexistence of the two theories allows different
perspectives which can only enhance our understanding of the
underlying physics. Secondly general relativity has failed to
explain properly a number of phenomena, see for instance
\cite{Flandern98}, while people keep trying to merge it with the
other main physical theory, the standard model. This suggests
that there may be some other theory that unifies everything, the
most successful so far being string theory \cite{Greene99}.

The field is wide open for new ideas that combine the best of
general relativity and the standard model, proving capable of
predictions as accurate as those provided by these two theories
but somehow having an entirely different approach. One must always
keep in mind that any model or theory is just a representation of
reality and not reality itself, the latter being rather difficult
to define without resorting to the observer's own perception. In
the present paper the author adopts an approach to his theory
based on the K-calculus by Herman Bondi \cite{Bondi80}, which has
been so successful in introducing relativity to undergraduates
all over the world. In his previous works the author used a
different approach and he hopes that this effort may render the
theory more acceptable to those that have difficulty accepting
that there may be alternatives to relativity. Furthermore there
issues related to simultaneity which have not been addressed so
far and are clarified in the present work.

The coordinate system used in 4-dimensional optics, henceforth
referred to as 4DO, differs from the relativistic space-time in
complementing the 3 spatial coordinates ($x^1 = x$, $x^2 = y$,
$x^3 =z$) with a 0th coordinate ($x^0 = \tau$) such that time
becomes a measure of geodesic arc length in Euclidean space.
Using the usual simplification of making $c =1$,
\begin{equation}
    \label{eq:arclength}
    \left(\mathrm{d}t \right)^2 = \left(\mathrm{d} \tau \right)^2
     + \left(\mathrm{d}x \right)^2 +
    \left(\mathrm{d}y \right)^2 + \left(\mathrm{d}z \right)^2.
\end{equation}
This equation is valid on tangent space only; for the more general
situation of curved space the definition is extended as
\begin{equation}
    \label{eq:arclength2}
    \left(\mathrm{d}t \right)^2 = g_{\alpha \beta}
    \mathrm{d}x^\alpha \mathrm{d}x^\beta,
\end{equation}
with $g_{\alpha \beta}$ the movement metric, rather than the
space metric as in general relativity \cite{Almeida01:4,
Almeida01:5}.

Understanding the meaning of coordinate $\tau$ is crucial for
proper handling of the theory in the most general situations, so
it was felt necessary to develop the basic theory in a way
similar to what is usually used to introduce relativity in
undergraduate courses, thereby giving the reader a good means for
assessing the similarities and differences between the two
theories. The author chooses to follow an introductory path
closely similar to what was introduced by Bondi \cite{Bondi80} and
later used by other authors, namely Martin \cite{Martin88} and
D'Inverno \cite{Inverno96}. For an observer traveling at speed
$v$, Bondi introduced the $K$ factor
\begin{equation}
    \label{eq:kfactor}
    K = \sqrt{\frac{c+v}{c-v}}\, ,
\end{equation}
which has since been used to identify this approach as
\emph{K-calculus}.

It is useful to remember that in 4DO the speed as we see it, what
will be designated by 3-speed, is actually the 3-dimensional
component of a 4-vector with the magnitude of the speed of light.
This was first explained in an introductory work by the author
\cite{Almeida00:4} and later developed in the works cited
previously. This approach ensures that 3-speed is bound by the
speed of light without resorting to hyperbolic space, as is the
case in relativity. Besides, it has been shown that 4DO holds the
potential for explaining physical processes which relativity has
failed to account for satisfactorily.

\section {Coordinate definition}
Bondi resorts to imaginary radar pulses bouncing on a distant
observer and a clock for the definition of space and time
coordinates. The idea is that the observer fixed on the
laboratory frame investigates the space around him with the help
of radar pulses and a clock. When a radar pulse is sent the
observer records the send instant $\tau_1$ and when the echo is
received from some location in space the observer records the
receive instant $\tau_3$ and the direction angles $\theta$ and
$\phi$; the reflection instant $\tau_2$ is inaccessible to the
observer. The direction angles provide two of the position
coordinates; remembering that the speed of light is set to unity,
the third position coordinate is the distance to the reflecting
object
\begin{equation}
    \label{eq:range}
    r = \frac{\tau_3-\tau_1 }{2}\, .
\end{equation}
The time coordinate is naturally given by
\begin{equation}
    \label{eq:timecoord}
    t = \frac{\tau_3 + \tau_1}{2}\, .
\end{equation}

Conversely 4DO replaces the time coordinate with a coordinate
defined by the reflection instant
\begin{equation}
    \label{eq:taucoord}
    \tau = \tau_2.
\end{equation}

The locus of an observer's points in either space is said to
define the observer's worldline in that particular space. The
graphical representation of 4-dimensional spaces is difficult, so
it is usual to suppress two spatial dimensions in order to
represent a 2-dimensional section of the space or to suppress one
spatial dimension and project the resulting 3-dimensional section
on the plane as a perspective diagram; this paper uses mainly the
2-dimensional section representation.

Suppose now an observer moving with uniform velocity $v$ in the
laboratory frame. This observer's worldline is a straight line on
both systems and for convenience it can be assumed that it passes
through the coordinate origin. In 2 dimensions the observer's
worldline has an equation $x=vt$; using Eqs.\ (\ref{eq:range},
\ref{eq:timecoord}) one can write
\begin{equation}
    \label{eq:uniform}
    \tau_3 - \tau_1  = v \left(\tau_3 + \tau_1 \right),
\end{equation}
which leads to the relation
\begin{equation}
    \label{eq:Bondi}
    \tau_3 = \frac{1+v}{1-v}~\tau_1.
\end{equation}
Considering the K-factor definition, Eq.\ (\ref{eq:kfactor}), it
is
\begin{equation}
    \label{eq:Bondi1}
    \tau_3 =  \tau_1 K^2.
\end{equation}

\begin{figure}[htb]
    \centerline{\psfig{file=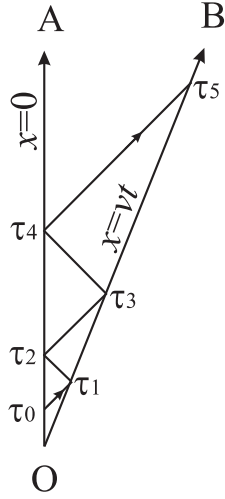, scale=0.6}}
    \caption{Minkowski space-time representation:
    Inertial observer B moves with relative velocity $v$ with respect
    to observer A. Observer A launches a radar pulse at instant $\tau_0$,
    which bounces back and forth between the two observers.}
    \label{fig:radar}
\end{figure}

Two inertial observers A and B are now considered, the latter
moving with velocity $v$ relative to A. Fig.\ \ref{fig:radar}
shows the Minkowski space-time worldlines of the two observers as
two straight lines diverging from the origin. The origin
placement at the point of intersection of the two worldlines does
not represent any limitation on the problem conditions. At instant
$\tau_0$ observer A launches a radar pulse which is bounced back
and forth successively between the two observers. The radar pulse
moves at the speed of light with respect to both observers so, in
observer A's frame its worldline is a broken line with alternating
segments of slope $+1$ and $-1$, respectively. The radar pulses
worldlines follow the general equation $t = \pm x + \tau_{2m-2}$,
with $m$ integer.

Observe that the $K$-factor is the same for both observers and so
it is possible to write Eq.\ (\ref{eq:Bondi}) generally as
\begin{equation}
    \label{eq:Bondi2}
    \tau_{n+1} = \frac{1+v}{1-v}~\tau_{n-1} = \tau_{n-1}K^2.
\end{equation}

The last equation defines recursively a succession of general term
\begin{equation}
    \label{eq:Bondi3}
    \tau_n = \tau_0 K^n.
\end{equation}

It is now possible to use Eq.\ (\ref{eq:Bondi3}) as an alternative
to Eq.\ (\ref{eq:taucoord}) for the definition of the $\tau$
coordinate
\begin{equation}
    \label{eq:taucoord2}
    \tau_n = \sqrt{\tau_{n+1} \tau_{n-1}}.
\end{equation}
Comparing with Eq.\ (\ref{eq:timecoord}) it becomes clear that
while Minkowski space-time defines time as the arithmetic mean
between the send and receive instants, 4DO uses a coordinate
defined as the geometric mean between the same instants for
$\tau$. This coordinate will be designated \emph{proper time} in
view of its relationship with the similarly designated quantity in
relativistic theory \cite{Almeida01:4}.

Eq.\ (\ref{eq:timecoord}) can be used to relate $t_n$ and $\tau_n$
eliminating $\tau_{n+1}$ and $\tau_{n-1}$ from the following three
equations:
\begin{displaymath}
    \tau_n = K \tau_{n-1},~~ \tau_{n+1} = K \tau_n,~~ t_n =
    \frac{\tau_{n+1} + \tau_{n-1}}{2},
\end{displaymath}
resulting in
\begin{equation}
    \label{eq:ttau}
    t_n = \gamma \tau_n,
\end{equation}
where
\begin{equation}
    \label{eq:gamma}
    \gamma = \frac{K + K^{-1}}{2}= \frac{1}{\sqrt{1 - v^2}}.
\end{equation}

\begin{figure}[htb]
    \centerline{\psfig{file=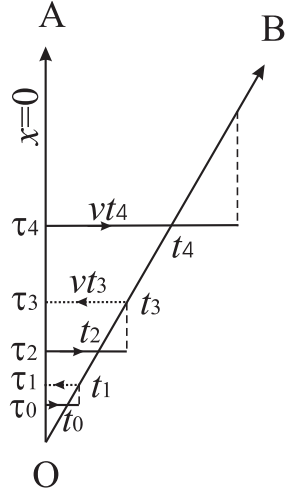, scale=0.6}}
    \caption{In optical space the observers move with velocity $c$ across
    4-dimensional space and time is measured along the worldline.}
    \label{fig:radaropt}
\end{figure}

Eqs.\ (\ref{eq:taucoord2} or \ref{eq:ttau}) allow the
construction of a graphic representation for the two observers'
4DO worldlines using coordinates $x, \tau$. Observer A's
worldline is drawn as the $\tau$ axis, because this observer does
not move in his own frame. For observer B it is sufficient to
plot the successive pairs $(x_n, \tau_n)$ and join them by a
straight line; B worldline's slope is $\tau_n/x_n$ and considering
Eq.\ (\ref{eq:ttau}) it is easily seen that this slope is given
by $\gamma/v$. Obviously light and radar pulses have $v=1$
\label{photons} and must be represented by horizontal lines of
zero slope.

The final graph is shown in Fig.\ \ref{fig:radaropt} but the
interaction between radar pulses and observers needs some
explanation. When the two observers leave the origin they have
synchronous clocks; time, however, is then evaluated separately
over each observer's worldline. When A launches the first radar
pulse, at instant $\tau_0$, the pulse is synchronous with A's
clock and evaluates time over it's own worldline. The pulse's
time is compared with B's own time at every point along the way;
only when they coincide can interaction occur and this happens
beyond the crossing point of the two worldlines on the graph. On
B's frame the interaction instant $t_1$ is translated into the
proper time $\tau_1$; this is represented by a dashed vertical
line joining the pulse's and B's worldlines.

The reflected pulse must suffer a similar process, only the roles
of observers A and B are now interchanged. The dashed lines
representing the reflected pulses at $\tau_1$ and $\tau_3$ must
not be mistaken by these pulses worldlines, because they are the
representation of worldlines on B's frame. On A's frame the
instants $\tau_1$ and $\tau_3$ are inaccessible and the pulses
actually return at $\tau_2$ and $\tau_4$, respectively.

From Eqs.\ (\ref{eq:ttau} and \ref{eq:gamma}) it is clear that $
\tau_n$, $v t_n$ and $ t_n$ are the sides of a rectangular
triangle. Remembering that $x_n = v t_n$, it is possible to write
\begin{equation}
    \label{eq:triangle}
    t^2 = \tau^2 + x^2.
\end{equation}
The subscripts have been suppressed because the relation most hold
for pulses sent out irrespective of the particular initial
instant $\tau_0$.

\begin{figure}[htb]
    \centerline{\psfig{file=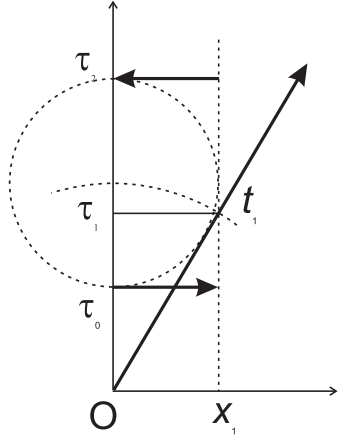, scale=0.6}}
    \caption{The figure shows a geometrical construction suitable
    for determining the worldline of a moving observer.}
    \label{fig:Construction}
\end{figure}

A simple geometrical construction can be used to plot the points
of the moving observer's worldline, Fig.\ \ref{fig:Construction}.
If a radar pulse is sent at $\tau_o$ and its reflection is
received at $\tau_2$, a circle is drawn with diameter
$\tau_2-\tau_0$ and, according to Eq.\ (\ref{eq:range}), the
moving observer is at a distance $x_1$ equal to the circle
radius. Considering Eq.\ (\ref{eq:timecoord}) the time $t_1$ for
the moving observer is equal to the distance from the origin to
the center of the circle. A new circle is now drawn with center
at the origin and radius equal to the value of $t_1$, in order to
transport this value to the moving observer's position. The value
of coordinate $\tau_1$ is automatically found by the intersection
of the second circle with the vertical line at position $x_1$.
The same process can then be repeated for all the pairs of sent,
received pulses.

\section{Interval \label{sec:radar}}
\begin{figure}[htb]
    \centerline{\scalebox{0.6}{\includegraphics{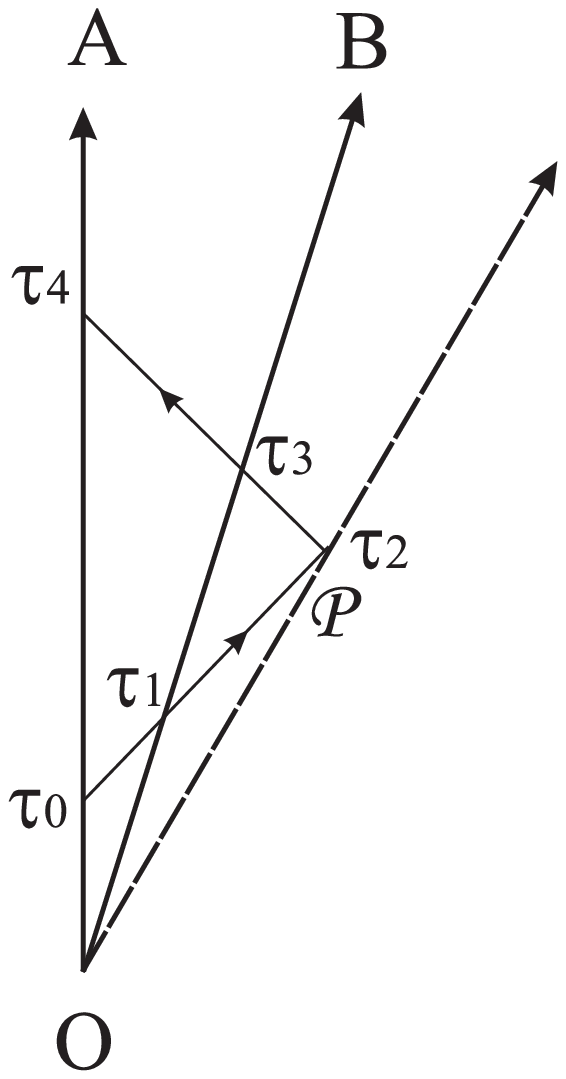}}
        \hspace{15mm} \scalebox{0.6}{\includegraphics{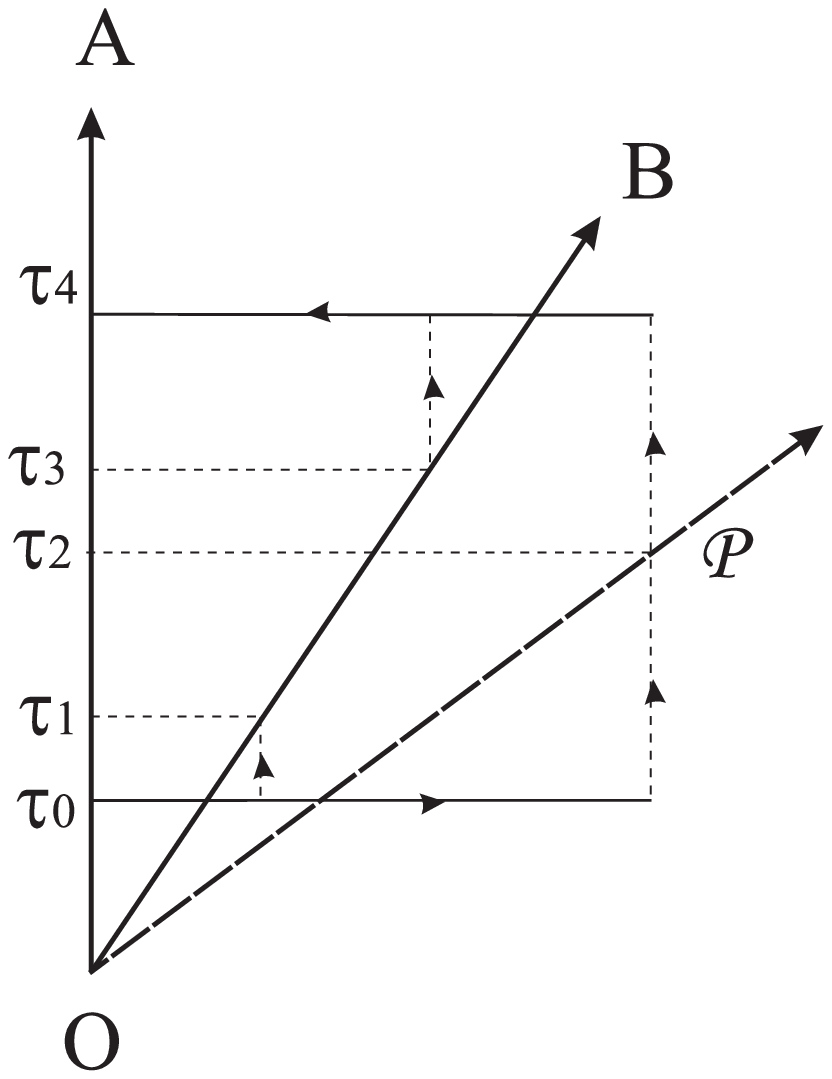}}}

    \centerline{ a) \hspace{45mm} b)}
    \caption{a) Minkowski space-time representation of two different observers
    looking at a third one ($\mathcal{P}$).
    A radar pulse is launched by observer A at instant $\tau_0$, passes B
    at instant $\tau_1$ and is reflected by $\mathcal{P}$ at instant $\tau_2$;
    on the way back it passes B at $\tau_3$ and arrives at A at instant
    $\tau_4$.
    b) The same situation in 4DO space.}
    \label{fig:interval}
\end{figure}

Consider now how two different inertial observes ''see'' a third
one. Fig.\ \ref{fig:interval} a)
is a Minkowski space-time representation of the worldlines of two
observers A and B, B moving with speed $v$ relative to A, and a
third observer $\mathcal{P}$,  moving with speed $u$ relative to
B. The first two observers use the radar formulas (\ref{eq:range},
\ref{eq:timecoord}, \ref{eq:taucoord2}) to compute
$\mathcal{P}$'s coordinates. Note first of all that
\begin{equation}
    \label{eq:twoobs2}
    \frac{\tau_1}{\tau_0} = K_v ,~~ \frac{\tau_2}{\tau_1} = K_u,~~
    \frac{\tau_3}{\tau_2} = K_u ,~~
    \frac{\tau_4}{\tau_3}= K_v.
\end{equation}
The coordinates of $\mathcal{P}$ evaluated by the two observers
are
\begin{equation}
    \label{eq:twoobs}
    \begin{array}{ll}
      x_A = (\tau_4-\tau_0)/2 ,~ & x_B = (\tau_3-\tau_1)/2, \\
      t_A = (\tau_4+\tau_0)/2,~ & t_B = (\tau_3+\tau_1)/2, \\
      \tau_A = \sqrt{\tau_0 \tau_4},~ & \tau_B = \sqrt{\tau_1
      \tau_3}~.
    \end{array}
\end{equation}
by direct replacement of $\tau_1$ and $\tau_3$ from Eqs.\
(\ref{eq:twoobs2}) it is possible to conclude that $\tau_A =
\tau_B =\tau_2$, i.\ e.\ the proper time coordinate is
independent of the inertial observer's own speed, as long as its
worldline passes through the origin. This conclusion can be seen
as equivalent to interval invariance in Minkowski space-time
 \cite{Martin88} \label{tau}.

The representation of Fig.\ \ref{fig:interval} b) shows how the
situation discussed above appears in 4DO space. The outgoing
pulse at $\tau_0$ is detected by observer B at $\tau_1$ and by
$\mathcal{P}$ at $\tau_2$. The return pulse is sent by
$\mathcal{P}$ at $\tau_2$ but is here represented by its
worldline in the laboratory frame, at $\tau_4$, unlike Fig.\
\ref{fig:radaropt}. Observer B detects the return pulse at
$\tau_3$.

It is possible to combine Eqs.\ (\ref{eq:twoobs2} and
\ref{eq:twoobs}) to obtain
\begin{equation}
    \label{eq:increments}
    t_A^2 = \tau_A^2 + x_A^2,~~
    t_B^2 = \tau_B^2 + x_B^2.
\end{equation}

Eqs.\ (\ref{eq:increments}) mean that the quantity $t^2$ can be
evaluated as $\tau^2 + x^2$ by any observer whose worldline
passes by the origin. If the problem is given a more general
formulation, in 4 dimensions and without the restriction of the
worldlines having to cross at the origin, it can be said that
\begin{equation}
    \label{eq:metricopt}
    \left(\mathrm{d}t \right)^2 = \delta_{\alpha \beta}
    \mathrm{d}x^\alpha \mathrm{d}x^\beta,
\end{equation}
which is equivalent to saying that \emph{time is a measure of
geodesic arc length in Euclidean \textnormal{4DO} space}.

The important results are then: 4DO space is Euclidean, with time
intervals being measured over the worldline of the observer. In a
change of coordinates from one observer to a different one, which
is in motion relative to the former, time intervals are not
preserved but rather proper time intervals.

In Minkowski space-time it is well known that the interval is
preserved in coordinate transformations and is given by
 \cite{Martin88, Inverno96}
\begin{equation}
    \label{eq:intervalmink}
    \left(\mathrm{d} \tau \right)^2 = \eta_{\bar{\alpha} \bar{\beta}}
    \mathrm{d}x^{\bar{\alpha}} \mathrm{d}x^{\bar{\beta}},
\end{equation}
with
\begin{equation}
    \label{eq:metricmink}
    \begin{array}{ll}
      \eta_{\bar{\alpha}\bar{\beta}}=1,
      & \bar{\alpha}=\bar{\beta}=0; \\
      \eta_{\bar{\alpha} \bar{\beta}}=-1,
      & \bar{\alpha}=\bar{\beta}\neq 0; \\
      \eta_{\bar{\alpha} \bar{\beta}}=0,
      & \bar{\alpha}\neq \bar{\beta}. \
    \end{array}
\end{equation}
for the set of coordinates
\begin{equation}
    \label{eq:coordmink}
    x^0 = t, ~~ x^1 = x, ~~ x^2 = y, ~~ x^3 = z.
\end{equation}
The bar over the indices is here used to designate Minkowski
space-time coordinates.

Eqs\ (\ref{eq:increments} and \ref{eq:intervalmink}) provide the
rules for elementary displacement transformation when changing
from one space to the other. Assume that the vector
$\mathrm{d}x^{\bar{\alpha}}$ is expressed in Minkowski space-time
and must be transformed to 4DO space; the transformation rule is
\begin{equation}
    \label{eq:minkopt}
    \begin{array}{ll}
      \mathrm{d}x^\alpha = \sqrt{\eta_{\bar{\mu} \bar{\nu}}
    \mathrm{d} x^{\bar{\mu}} \mathrm{d} x^{\bar{\nu}}},
    & \alpha=0;  \\
      \mathrm{d} x^\alpha = \mathrm{d}x^{\bar{\mu}},
      & \alpha=\bar{\mu} \neq 0. \
    \end{array}
    \end{equation}
The opposite transformation follows a similar rule
\begin{equation}
    \label{eq:optmink}
    \begin{array}{ll}
      \mathrm{d}x^{\bar{\alpha}} = \sqrt{\delta_{\mu \nu}
    \mathrm{d} x^\mu \mathrm{d} x^\nu},
    & \bar{\alpha}=0;  \\
      \mathrm{d} x^{\bar{\alpha}} = \mathrm{d}x^\mu,
      & \bar{\alpha}=\mu \neq 0. \
    \end{array}
\end{equation}

It is possible to define the 4-velocity for an observer in the two
spaces as $\dot{x}^\alpha = \mathrm{d}x^\alpha /\mathrm{d}t$ and
$\dot{x}^{\bar{\alpha}}= \mathrm{d}x^{\bar{\alpha}}
/\mathrm{d}t$, respectively, with the $t$ derivatives taken over
the worldline. The 4-velocity transformation between the two
spaces follows rules similar to those expressed by Eqs.\
(\ref{eq:minkopt}, \ref{eq:optmink}) with
$\mathrm{d}x^{\bar{\alpha}}$ and $\mathrm{d}x^\alpha$ replaced by
$\dot{x}^{\bar{\alpha}}$ and $x^\alpha$.

\section{\label{Lorentz}Lorentz equivalent transformations}
The subject of this paragraph has already been considered in
 \cite{Almeida01:4} but needs to be reviewed here for completeness.
Our approach to coordinate transformation between observers moving
relative to each other is different from special relativity; while
in the latter case the interval is given by Eq.\
(\ref{eq:intervalmink}), thus ensuring that a coordinate
transformation preserves the interval and affects both spatial
and time coordinates, our option of making time intervals measure
geodesic arc length gives time a meaning independent of any
coordinate transformation. We thus propose that Lorentz
equivalent transformations between a "fixed" or "laboratory"
frame and a moving frame are a combination of two simultaneous
processes\label{twoproc}. The first process is a tensorial
coordinate transformation, which changes the coordinates keeping
the origin fixed, with no influence in the way time is measured,
while the second process corresponds to a "jump" into the moving
frame, changing the metric but not the coordinates.

It has been established in page \pageref{photons} that photons
follow worldlines of $\mathrm{d}\tau=0$ and carry the value of the
$\tau$ coordinate across the Universe, which was confirmed in
page \pageref{tau} by the verification that proper time intervals
were independent of the observer. On the other hand the time
interval must also evaluate to the same value on all coordinate
systems of the same frame, due to its definition as interval of
4DO space.

\begin{figure}[thb]
    \centerline{\psfig{file=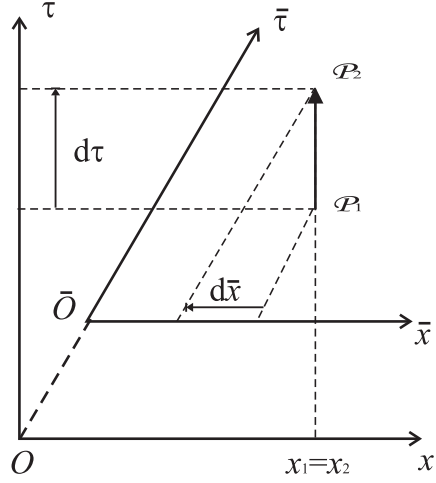, scale=0.7}}
    \caption{\label{fig:Lorentz}The worldline of moving observer $\bar{O}$
    coincides with the
    $\bar{\tau}$ or $x^{\bar{0}}$ axis, while the $\bar{x}$
    axis stays parallel to the $x$ axis.
    A displacement parallel to axis $\tau$ or $x^0$ implies
    both $\bar{x}$ and $\bar{\tau}$
    components in the moving frame.}
\end{figure}

One can consider observers $O$ and $\bar{O}$, the latter moving
along one geodesic of $O$'s coordinate system, as depicted in
Fig.\ \ref{fig:Lorentz}. Let the geodesic equation have a
parametric equation $x^{\alpha'} (t)$. The aim is to to find a
coordinate transformation tensor between the two observers'
coordinate systems, ${\Lambda^{\bar{\mu}}}_\alpha =
\partial x^{\bar{\mu}} / \partial x^\alpha$; it has already
established that $\mathrm{d} x^{\bar{0}} = \mathrm{d} x^0$ and so
it must be ${\Lambda^{\bar{0}}}_0 = 1$ and ${\Lambda^{\bar{0}}}_i
= 0$.

A photon traveling parallel to the $x^1$ direction follows a
geodesic characterized by $\mathrm{d}t = \mathrm{d} x^1$ on $O$'s
coordinates. On $\bar{O}$'s coordinates the photon will move
parallel to $x^{\bar{1}}$ and so it must be also $\mathrm{d}t =
\mathrm{d} x^{\bar{1}}$. The same behaviour could be established
for all three $(x^i,~x^{\bar{i}})$ coordinate pairs and one
concludes that ${\Lambda^i}_{\bar{i}} = 1$.

Consider now two points $\mathcal{P}_1$ and $\mathcal{P}_2$ on a
line parallel to $x^0$ axis in $O$'s coordinates, separated by a
time interval $\mathrm{d}t = \mathrm{d} x^0$. On $\bar{O}$'s
coordinates it will be
\begin{equation}
    \mathrm{d} x^{\bar{i}} = -\dot{x}^{i'} \mathrm{d}t
    =-\frac{\dot{x}^{i'}}{\dot{x}^{0'}}~\mathrm{d} x^0,
\end{equation}
allowing the conclusion that ${\Lambda^{\bar{i}}}_0 =
-\dot{x}^{i'}/\dot{x}^{0'}=-\breve{x}^i$, where the notation
$\breve{x}^i$ is used for derivation with respect to $\tau$. Fig.\
\ref{fig:Lorentz} shows graphically the relation between the two
coordinate systems.

The coordinate transformation tensor between $x^\alpha$ and
$x^{\bar{\mu}}$ is consequently defined as
\begin{equation}
    \label{eq:transform}
    {\Lambda^{\bar{\mu}}}_\alpha = \left[\begin{array}{cccc}
      1 & 0 & 0 & 0 \\
      -\breve{x}^{1'} & 1 & 0 & 0 \\
      -\breve{x}^{2'} & 0 & 1 & 0 \\
      -\breve{x}^{3'} & 0 & 0 & 1 \
    \end{array} \right].
\end{equation}
The reverse transformation is obtained changing the sign of
$-\breve{x}^i$.

The metric for $\bar{O}$'s coordinates can now be evaluated
\begin{equation}
    \label{eq:movmetr}
    g_{\bar{\mu} \bar{\nu}} = {\Lambda^{\alpha}}_{\bar{\mu}}
    {\Lambda^{\beta}}_{\bar{\nu}} g_{\alpha \beta};
\end{equation}
and after substitution it is
\begin{equation}
    \label{eq:movmetr2}
    g_{\bar{\mu} \bar{\nu}} =
    \left[\begin{array}{cccc}
      1 + \delta_{a' b'}\breve{x}^{a'}\breve{x}^{b'}
      & \breve{x}^{1'}
      & \breve{x}^{2'} & \breve{x}^{3'} \\
      \breve{x}^{1'} & 1 & 0 & 0 \\
      \breve{x}^{2'} & 0 & 1 & 0 \\
      \breve{x}^{3'} & 0 & 0 & 1 \
    \end{array}\right].
\end{equation}

The metric given by the previous equation will evaluate time in
$\bar{O}$'s coordinates in a frame which is fixed relative to
$O$, that is, this is still time as measured by observer $O$.
Observer $\bar{O}$ will measure time intervals
$\mathrm{d}\bar{t}$ in his own frame and so, although the
coordinates are the same as they were in $O$'s frame, time is
evaluated with the Kronecker metric $\delta_{\bar{\mu} \bar{\nu}}$
instead of metric given by Eq.\ (\ref{eq:movmetr2}).
Mathematically the metric replacement corresponds to the identity
\begin{equation}
    \label{eq:identity}
    \delta_{\bar{\alpha} \bar{\beta}}
    =\left(g_{\bar{\mu}\bar{\nu}}\right)^{-1}
     g_{\bar{\mu}\bar{\nu}} .
\end{equation}

It was said in page \pageref{twoproc} that the transition from
the laboratory frame to a moving observer's frame was the result
of two separate processes, one of which was responsible for the
change of coordinates and the other one for the time change. The
first process can now be identified with a coordinate change
governed by the tensor ${\Lambda^{\bar{\mu}}}_\alpha$, while the
second one involves multiplication of the metric by
$g^{\bar{\mu}\bar{\nu}}$. The second process is intrinsically
separate from the first one and was used in \cite{Almeida01:4,
Almeida01:5} to derive the Lorentz force due to a moving charge
and the center of mass equivalent of an orbiting mass.

Note that this double process preserves the essential relationship
of special relativity, i.\ e. $\mathrm{d}\bar{t}^2 -
\delta_{\bar{i}
\bar{j}}\mathrm{d}x^{\bar{i}}\mathrm{d}x^{\bar{j}} =
\mathrm{d}t^2-\delta_{ij} \mathrm{d} x^i \mathrm{d} x^j$.
\subsection{Time contraction for a moving observer}
Eq.\ (\ref{eq:increments}) established that time intervals should
be measured by the arc length of an observer's worldline.
Recalling Fig.\ \ref{fig:Lorentz}, the time measured for observer
$\bar{O}$'s trajectory in the laboratory frame of observer $O$ is
given by $\mathrm{d}t = \gamma \mathrm{d}\tau$, according to Eq.\
(\ref{eq:ttau}). Observer $\bar{O}$ is at rest in his own frame
and sees his frame axes normal to each other, i.e.\ he uses the
Kronecker metric, and so he measures for himself a time interval
$\mathrm{d}\bar{t} = \mathrm{d}\tau$. Obviously
$\mathrm{d}\bar{t}< \mathrm{d}t$ and so we say that the time
measured by a moving observer runs slower than the time measured
on the laboratory frame.

At this point it is useful to give physical meaning to the
coordinate $\tau$, so that 4DO is understood as a space with
physical significance. This space is designed to represent the
movement of observers through 3-dimensional space. Accordingly it
uses the standard coordinates to identify the observers'
positions, eventually scaled by the observers' masses as
described in Refs.\ \cite{Almeida01:4, Almeida01:5}. The 0th
coordinate $\tau$ does not have a Universal meaning, as was shown
in the previous paragraph; it is just the time indicated by the
observers' own clocks and so it does deserve to be called proper
time. The fact that coordinate $\tau$ can be identified with each
observer's own time is of no consequence for the representation
of his own worldline but becomes important when two observers
interact, for then time must evaluate to the same value when both
observers' worldlines are represented in a common frame. The
problems of interaction between observers and the concept of
simultaneity will be dealt with in paragraph \ref{collision}.
\subsection{The twin paradox in optical space}
It is customary to discuss the case of two twins $A$ and $B$, of
which $A$ stays at home, remaining inertial at all times, while
$B$ goes on an extended space flight at high speed before he
eventually joins $A$ back home. The reasoning is that as $B$'s
time is constantly running slower than $A$'s, $B$ must come back
younger than $A$.

In 4DO the argument is inconsistent because one must distinguish
between the time measurements on the frames of the two twins. In
fact $B$'s time measured on his own frame is shorter than his time
measured by $A$ on the fixed frame. On the other hand $B$'s time
measured by himself is exactly the same as $A$'s time measured on
the fixed frame and the two twins actually grow old together. A
proper discussion of this problem should not ignore the fact that
$B$'s frame is necessarily accelerated and this has also an
influence on the way clocks run.

\section{Lagrangean definitions for geodesics in Minkowski and 4DO}
If $\sigma$ is a parameter along a geodesic, for any space the
geodesic equations can be derived from the Lagrangean
\cite{Martin88}
\begin{equation}
    \label{eq:lagrangean}
    L\left(x^\alpha,\frac{\mathrm{d}x^\alpha}{\mathrm{d}\sigma}\right)
    = g_{\alpha \beta}
    \frac{\mathrm{d}x^\alpha}{\mathrm{d}\sigma} \frac{\mathrm{d}x^\beta}
    {\mathrm{d}\sigma}.
\end{equation}
The geodesic equations are then
\begin{equation}
    \label{eq:geodesic}
    \frac{\mathrm{d}}{\mathrm{d} \sigma} \left(\frac{1}{\sqrt{L}}
    \frac{\partial L}
    {\displaystyle{\partial \frac{\mathrm{d}x^\alpha}{\mathrm{d}\sigma}}}
     \right)
    -\frac{1}{\sqrt{L}}
    \frac{\partial L}{\partial x^\alpha}=0,
\end{equation}
with $n-1$ independent equations on an $n$-dimensional space. If
the geodesic arc length is chosen as parameter the Lagrangean
becomes unity.

It can be shown that the geodesic equations are equivalent in
Minkowski and optical spaces. For this one chooses arc length as
parameter in both spaces and derives the geodesic equations
starting from the Lagrangean.

Considering Eq.\ (\ref{eq:intervalmink}), in 2-dimensional
Minkowski space-time the Lagrangean is
\begin{equation}
    \label{eq:lagmink}
    L_\mathrm{M}=\left(\frac{\mathrm{d}t}{\mathrm{d}\tau}\right)^2
    - \left(\frac{\mathrm{d}x}{\mathrm{d}\tau}\right)^2 =1.
\end{equation}
As the Lagrangean is independent of $t$ one can write
\begin{equation}
    \label{eq:partialmink}
    \frac{\partial L_\mathrm{M}}{\partial \displaystyle{\frac{\mathrm{d}t}
    {\mathrm{d}\tau}}}= 2\frac{\mathrm{d}t}{\mathrm{d}\tau}
    =\mathrm{constant}.
\end{equation}

Making the second member of the equation above equal to $2\gamma$
and replacing in Eq.\ (\ref{eq:lagmink})

\begin{equation}
    \label{eq:lag1}
    1 = \gamma^2 -\left(\frac{\mathrm{d}x}{\mathrm{d}\tau} \right)^2.
\end{equation}
Noting that $\mathrm{d}x/\mathrm{d}\tau =\gamma
\mathrm{d}x/\mathrm{d}t = \gamma v$
\begin{equation}
    \label{eq:lag2}
    1 = \gamma^2 \left(1 - v^2 \right),
\end{equation}
which is an equation equivalent to Eq.\ (\ref{eq:gamma}).

Considering now a 2-dimensional section of 4DO space and using
again arc length as parameter, the Lagrangean is
\begin{equation}
    \label{eq:lagopt}
    L_\mathrm{O}=\left(\frac{\mathrm{d}\tau}{\mathrm{d} t} \right)^2
     + \left(\frac{\mathrm{d} x}{\mathrm{d} t} \right)^2 = 1.
\end{equation}
The fact that the Lagrangean does not depend on $\tau$ allows one
to set $\mathrm{d} \tau/\mathrm{d}t= 1/\gamma$ and replace above
to get
\begin{equation}
    \label{eq:lagopt2}
    \frac{1}{\gamma^2} + v^2 =1,
\end{equation}
the same as Eq.\ (\ref{eq:lag2}).

Eqs.\ (\ref{eq:lag1}, \ref{eq:lag2}) were both derived from the
Lagrangeans and state the basic relationship between time and
proper time that had been established before. It is then clear
that both spaces are equally adequate for derivation of the
geodesic equations.

Although the geodesic equations are the same in both spaces the
Lagrangeans are not equivalent; when $t$ is used as parameter in
Eq.\ (\ref{eq:lagrangean}) the Minkowski Lagrangean simplifies to
$L_\mathrm{M}= 1 - v^2$ while the 4DO Lagrangean is still given
by Eq.\ (\ref{eq:lagopt}). The implications of this difference
will be discussed below.
\section{Hamiltonian and particle energy}
The Hamiltonian of a system is a function of coordinates,
conjugate momenta and a parameter, that describes the system
evolution in phase space. Phase space associated with
4-dimensional coordinate space should then be 9-dimensional. In
classical mechanics one is usually led to use time as parameter,
thus eliminating two dimensions in phase space. What follows
compares phase spaces associated with 2-dimensional sections of
Minkowski and 4DO spaces.

It has been established that the Lagrangean must be unity when
expressed in terms of the arc length so the action in Minkowski
space-time is
\begin{equation}
    \label{eq:actionmink}
    \delta \int \mathrm{d}s = \delta \int
    \sqrt{(\mathrm{d}t)^2 -(\mathrm{d} x)^2}
    = \delta \int \sqrt{1 - \dot{x}^2}~ \mathrm{d}t =0,
\end{equation}
with ''dot'' used to represent time derivative. The Lagrangean
with time as parameter can then be taken as \cite{Goldstein80}
\begin{equation}
    \label{eq:lagmink2}
    L_\mathrm{M}=\sqrt{1 - \dot{x}^2}=\frac{1}{\gamma}~,
\end{equation}
from which it is possible to derive the conjugate momentum
\begin{equation}
    \label{eq:momentmink}
    p_\mathrm{M} = \frac{\partial L_\mathrm{M}}
    {\partial \dot{x}}= -\gamma \dot{x},
\end{equation}
and the Hamiltonian
\begin{equation}
    \label{eq:hamiltmink}
    H_\mathrm{M}= p_\mathrm{M} \dot{x} - L_\mathrm{M}= -\frac{1}{\gamma}.
\end{equation}

Eq.\ (\ref{eq:hamiltmink}) justifies the statement that a unit
mass body's energy is given by $-1/\gamma$ and quite naturally, if
the body has mass $m $ one says that the energy increases
proportionally.

Looking at 4DO space one can perform a similar derivation; the
action is now
\begin{equation}
    \label{eq:actionopt}
    \delta \int \mathrm{d}t = \delta \int \sqrt{(\mathrm{d}
    \tau)^2 + (\mathrm{d}x)^2}
    = \delta \int { \sqrt{1 +  \left(\frac{\mathrm{d}x}
    {\mathrm{d} \tau}\right)^2}~ \mathrm{d}\tau} =0.
\end{equation}
from which the Lagrangean and conjugate momentum with $\tau$ as
parameter can be derived
\begin{eqnarray}
    L_\mathrm{O}&=&\gamma
    =\sqrt{1 + \left(\frac{dx}{d \tau}\right)^2}, \\
    p_\mathrm{O} &=& \frac{\partial L_\mathrm{O}}{\partial \tau} =
    \frac{1}{\gamma}\frac{\partial x}{\partial \tau}.
\end{eqnarray}
Note that $p_\mathrm{O} = \dot{x}/(\gamma \dot{\tau}) $ and
considering Eq.\ (\ref{eq:ttau})
    $p_\mathrm{O} = \dot{x}$.

The Hamiltonian becomes
\begin{equation}
    \label{eq:hamiltopt}
    H_\mathrm{O}= p_\mathrm{O} \frac{\partial x}{\partial \tau}
     - L_\mathrm{O}= -\frac{1}{\gamma}.
\end{equation}

It is tempting to see the fact that $H_\mathrm{M}=H_\mathrm{O}$
as a confirmation that both spaces are equivalent for the study
of a body's movement. This is actually not quite true!

The differences arise when the body is influenced by a coordinate
dependent potential $V$, in which case one should write $H = -m
/\gamma +V$; in Minkowski space one gets the familiar canonical
equation
\begin{equation}
    \label{eq:canonmink}
    \frac{\mathrm{d} p_\mathrm{M}}{\mathrm{d} t}
    = - \frac{\partial H}{\partial x},
\end{equation}
while in 4DO space it is
\begin{equation}
    \label{eq:canonopt}
    \frac{\mathrm{d} p_\mathrm{O}}{\mathrm{d} \tau} = \gamma
    \frac{\mathrm{d}}{\mathrm{d} t}\left(\frac{p_\mathrm{M}}{\gamma}\right)
    =- \frac{\partial H}{\partial x}.
\end{equation}

The difference between Eqs.\ (\ref{eq:canonmink}) and
(\ref{eq:canonopt}) exists only when $\gamma$ is time dependent;
this happens if the velocity is varying not only in direction but
also in magnitude. In fact the problem only arises because a
potential was considered but it has been stated already
\cite{Almeida01:4} that 4DO precludes potentials and considers
all interactions to be dealt with through space curvature and
associated metric modification. The problem is then ill-posed;
nevertheless the predictions of general relativity will not be
coincident with 4DO's whenever the former theory appeals to
potentials.
\section{Conservation law and simultaneity}
\label{collision}
Understanding the physical meaning of proper time is important if
one wishes to acquire the feeling of the physical phenomena that
are expressed in optical space. One is used to the concept of
simultaneity and the discussion in section \ref{sec:radar} has
already shown that this concept is not easily translated into the
new space; the following lines discuss a typical problem involving
simultaneity, which hopefully will clarify this point. A
collision situation is as good as one can get to discuss
simultaneity, with the bonus that it will  bring conservation
into the picture.

First of all it will be admitted that for some fortunate reason
the coordinate origin coincides exactly with the collision point,
so that the simultaneity problem is solved automatically. Two
particles are to be considered, $a$ and $b$, with position vectors
in 2D optical space given by
\begin{equation}
    \label{eq:posvec}
    \mathbf{r}_a = \left(\tau_a,x_a \right),~~
    \mathbf{r}_b = \left(\tau_b,x_b \right).
\end{equation}
By derivation with respect to $t$ it is possible to define the
speed vectors
\begin{equation}
    \label{eq:speedvec}
    \mathbf{v}_a = \left(\frac{1}{\gamma}_a, v_a \right),~~
    \mathbf{v}_b = \left(\frac{1}{\gamma}_b, v_b \right);
\end{equation}
both vectors have unit length. The momentum vectors are
\begin{equation}
    \label{eq:momentvec}
    \mathbf{p}_a = m_a\left(\frac{1}{\gamma}_a, v_a \right),~~
    \mathbf{p}_b = m_b\left(\frac{1}{\gamma}_b, v_b \right),
\end{equation}
with $m_a$, $m_b$ the masses of particles $a$ and $b$,
respectively. Notice that coordinate scaling of the particles'
coordinates by their respective masses cannot be used in
graphical representations, although it is useful for mathematical
calculations, as explained in Refs.\ \cite{Almeida01:4,
Almeida01:5}.

Momentum conservation implies that at any position along $x$ the
sum of the two momenta must be preserved; in particular at $x=0$
one can write the two equations
\begin{equation}
    \label{eq:conserven}
    \frac{m_a}{\gamma_a} + \frac{m_b}{\gamma_b}
    = \frac{m_a}{\gamma'_a} + \frac{m_b}{\gamma'_b},
\end{equation}
\begin{equation}
    \label{eq:conserm}
    m_a v_a + m_b v_b = m_a v'_a + m_b v'_b,
\end{equation}
where the prime is used to denote the values after the collision.

The first of these equations is obviously an energy conservation
law when Eq.\ (\ref{eq:hamiltopt}) is considered and can be
written
\begin{equation}
    \label{eq:conserven2}
    m_a \sqrt{1-v_a^2} + m_b \sqrt{1-v_b^2}
    = m_a \sqrt{1-{v'_a}^2} + m_b \sqrt{1-{v'_b}^2}.
\end{equation}

\begin{figure}[htb]
    \centerline{\psfig{file=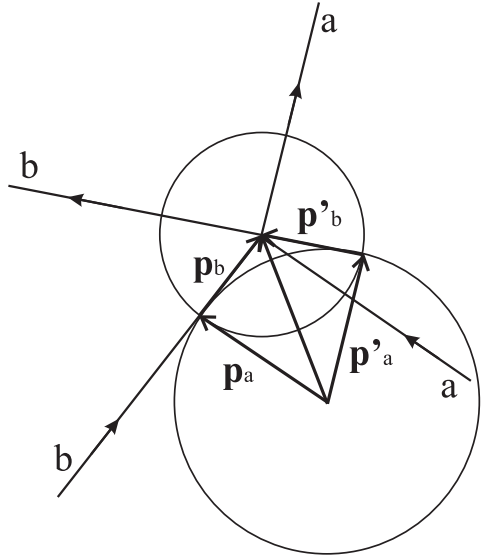, scale=0.6}}
    \caption{Head on collision of two particles. Conservation of 2D
    optical space momentum determines both particle speed after collision.}
    \label{fig:collision}
\end{figure}
The length of the particles momenta stay unaltered by the
collision; only their directions are changed. Fig.\
(\ref{fig:collision}) exemplifies a collision situation. The
addition of the momenta $\mathbf{p}_a + \mathbf{p}_b$ gives the
total momentum that must stay unaltered by the collision. The
length of individual momenta must also be preserved because the
particles' masses don't change. The construction shows the only
other way that two vectors of lengths $|\mathbf{p}_a|$ and
$|\mathbf{p}_b|$ can be added so that their sum is preserved;
these vectors are $\mathbf{p}'_a$ and $\mathbf{p}'_b$,
respectively and their orientation gives the speed of both
particles after the collision.

\begin{figure}[htb]
    \centerline{\psfig{file=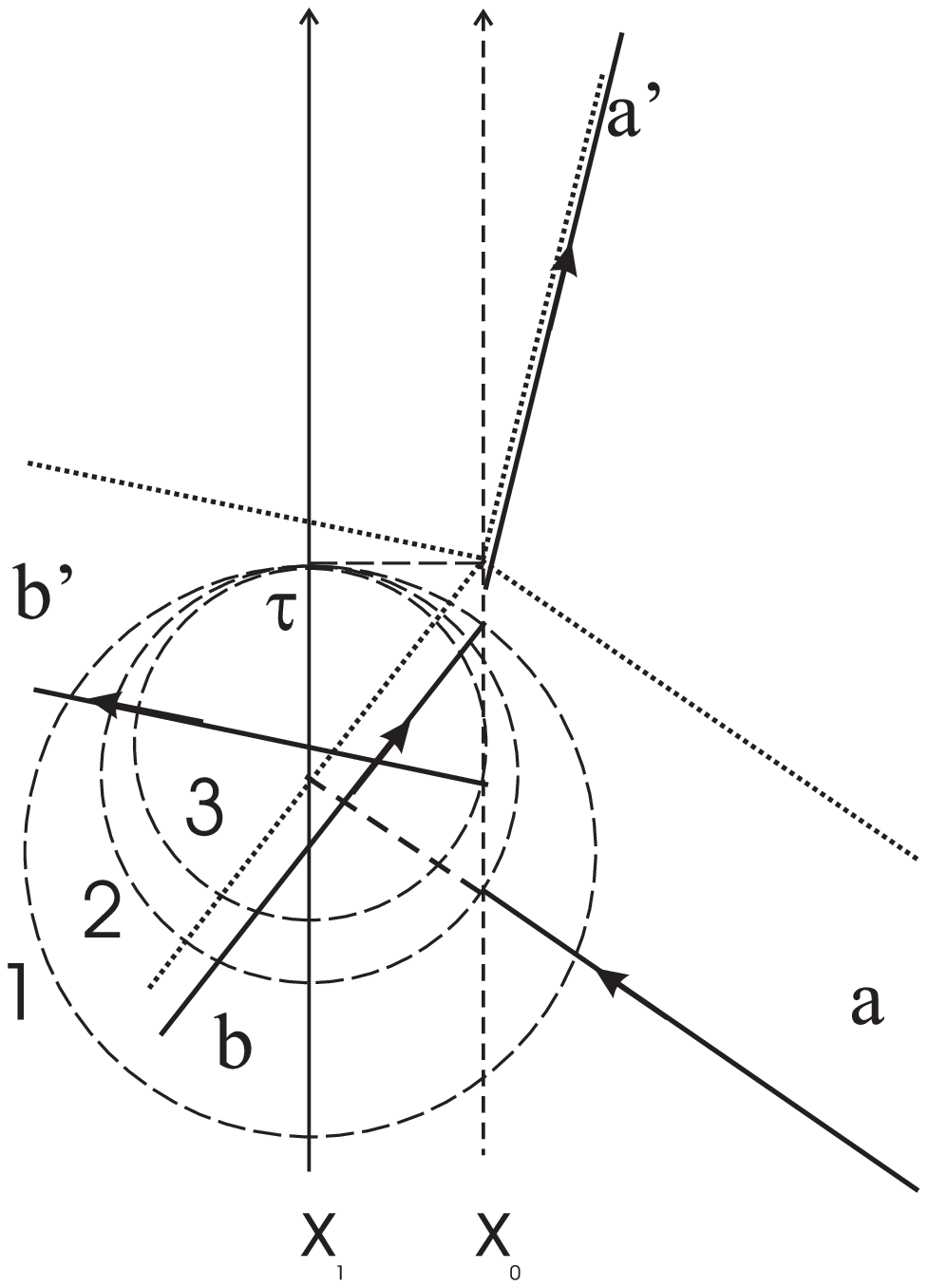, scale=0.6}}
    \caption{When represented on a distant observer's frame the
    worldlines no longer intersect at the collision position.}
    \label{fig:collision2}
\end{figure}

The problem was analyzed from the point of view of an observer
standing at the collision position. When represented on a distant
observer's frame, the particles' worldlines no longer intersect
at the collision position, as shown in Fig.\ \ref{fig:collision2}
where $x_0$ and $x_1$ are the collision and observer's positions,
respectively. The collision instant $\tau$ is the time measured
along each of the four different worldlines. The geometrical
construction on the figure is such that circles 1, 2 and 3 have
centers on the intersection of the observer's worldline, at $x_1$
with the worldlines $b$, $a$ and $b'$, respectively. The circle
associated with $a'$ worldline is not shown. All the circles
cross $x_1$ at a common point $\tau$, ensuring that each circle's
radius is exactly the time that mediates between the instant of
the associated worldline passing at $x_1$ and the collision
instant. The point of intersection of each circle with the
vertical line at $x_0$ marks the point of passage of the
associated worldline by this position. The dotted lines meeting
at $x_0$ represent the situation depicted on Fig.\
\ref{fig:collision} and clearly highlight the difficulty in
representing simultaneous events on distant observers' frames.
\section{Conclusions}
4DO has been shown in previous work to be a viable alternative to
general relativity, being able to produce identical predictions
in most observable situations but offering reasonable
explanations for phenomena that Einstein's theory has not been
able to accommodate comfortably. Furthermore, 4DO shows prospects
of being compatible with standard theory, paving the ground for a
unified theory of Physics, which has so far been the realm of
string theory.

4DO is based on the use 4-dimensional movement space, resulting
from the consideration of the usual 3-dimensional coordinates
complemented by proper time. The physical meaning of this space
is not intuitive and the present paper uses the established
K-calculus to make a parallel derivation of special relativity and
4DO, allowing a real possibility of comparison between the two
theories. The significance of the proper time coordinate is given
special attention and its definition is made very clear in terms
of just send and receive instants of radar pulses.

Special relativity and 4DO are also compared in terms of
Lagrangian definition of worldlines and movement Hamiltonian. The
final section of the paper discusses simultaneity through the
solution of a two particle head-on collision problem. It is shown
that a very simple graphical construction automatically solves
energy and momentum conservation when the observer is located at
the collision position. A further discussion of the
representation for a distant observer further clarifies how
simultaneity is accommodated by 4DO.
  \bibliography{aberrations}   
  \bibliographystyle{unsrt}

\end{document}